%% file: YULE_AML.tex
\documentclass{amsart}
\usepackage{epsfig}
\usepackage{latexsym}

\newtheorem{theorem}{Theorem}

\newtheorem{corollary}[theorem]{Corollary}

\usepackage{geometry} % see geometry.pdf on how to lay out the page. There's lots.
\usepackage{amsmath}
\newcommand{\PP}{{\mathbb P}}
\newcommand{\EE}{{\mathbb E}}

\begin{document}

\title[Edge lengths in a Yule tree]{Expected length of pendant and interior edges of a Yule tree}
\author{Mike Steel and Arne Mooers}
\address{MS: Allan Wilson Centre for Molecular Ecology and Evolution, Biomathematics Research
Centre, University of Canterbury, Christchurch, New Zealand; AOM: IRMACS, Simon Fraser University, Burnaby, Canada.}

\email{m.steel@math.canterbury.ac.nz}

\begin{abstract}
The Yule (pure-birth) model is the simplest null model of speciation;  each lineage gives rise to a new lineage independently with the same rate $\lambda$. We investigate the 
expected  length of an edge chosen at random from the resulting evolutionary tree. In particular, we compare the expected length of a randomly selected edge with the expected length of a randomly selected pendant edge. We provide some exact formulae, and show how our results depend slightly on  whether the depth of the tree or the number of leaves is conditioned on, and whether $\lambda$ is known or is estimated using maximum likelihood. 
\end{abstract}


\maketitle

\section{Introduction}
In evolutionary biology, the simplest model of speciation assumes that, at any moment, each of the then-extant lineages randomly gives rise to a new lineage at some constant rate (and independently of other such events).  This model, and an extension, was described by Yule in 1924  \cite{yule}. It generates a rooted binary tree for which each edge has an associated random length -- the duration of a lineage until it speciates (i.e. gives rise to a new lineage).  The Yule model is widely used in phylogenetic analysis; often, extinction is also allowed, but in this short note, we deal only with the pure-birth model. 

Although many properties of the Yule model have been extensively investigated over the years (e.g. \cite{ger2, Nee, yan}), in this paper we consider a question that has received less attention -- namely what can one say about the expected length of an edge selected uniformly at random from the set of pendant edges, or from all edges (pendant and interior)? 

We derive simple exact formulae for these quantities under two scenarios: either the number of leaves is given (but not the depth of the tree) or the depth of the tree is given (but not the number of leaves). We also evaluate these formulae when the diversification rate is replaced by its maximum likelihood estimate based on the depth of the tree and the number of leaves.  We will work with expected
average edge lengths  (these being the same as the expected edge length of an edge selected uniformly at random from the appropriate class of edges  - pendant or interior).

Consider then a pure-birth Yule tree with diversification rate $\lambda$.  The time that a given lineage persists until it speciates has an exponential distribution with a mean of $\frac{1}{\lambda}$.
We will assume throughout that the tree starts as an initial bifurcation -- that is, initially at some time $t$ in the past, it has two lineages each of length 0, as in \cite{Nee}.   If there are $k$ lineages present at a given moment, then the expected time until the next speciation event is  also exponentially distributed, and with a mean of $\frac{1}{k \lambda}$.   After time $t$  from the initial bifurcation, we produce a binary tree; the  expected number of leaves in the tree is $2e^{\lambda t}$.  

Since $\frac{1}{\lambda}$ is the expected time that a lineage persists until it speciates, it might be expected that the expected length of a randomly selected edge (pendant or interior) in a Yule tree would also be  $\frac{1}{\lambda}$. However, we will see that the expected value is either exactly or approximately equal to one-half this value, depending on what is being conditioned on.   The intuitive explanation for the $\frac{1}{2}$ factor is that we are considering expected edge lengths in a bifurcating (binary) tree, rather than a linear sequence of events. 

\section{Expected pendant vs. interior edge lengths as function (only) of $n$} 

In this section, we show that regardless of when we observe a tree with $n$ leaves, the expected length of a random interior edge length is $\frac{1}{2\lambda}$. For a  randomly chosen pendant edge  (the length of the branch leading from a species back to where it first meets the rest of the tree), the expected value depends on when it is observed, but it converges to $\frac{1}{2\lambda}$ as $n$ grows, and is exactly equal to $\frac{1}{2\lambda}$ under a null assumption concern the depth of the tree. 

Consider growing  the Yule tree from the initial bifurcation until it has $n+1$ leaves. Of course, the time ($t$) that this takes is a random variable, and we will suppose in this section that $t$ is not known.
Let $P_n$ be the expected value of sum of the lengths of the pendant edges of the tree on $n$ leaves up to (just before) we first get $n+1$ leaves, and
let $p_n = P_n/n$ be the expected value of the average pendant edge length. Similarly, let $I_n$ be the expected sum of the lengths of the interior edges up to (just before) we get $n+1$ leaves, and let $i_n = I_n/(n-2)$ be the expected value of the average interior edge length.
\begin{theorem}
\label{firstbasic}
For all $n \geq 3$, 
$i_n = p_n = \frac{1}{2\lambda}.$
\end{theorem}
{\em Proof:}   We have the following two recursions for $n \geq 3$:
\begin{equation}
\label{recurs1}
I_n = I_{n-1} + \frac{P_{n-1}}{n-1};
\end{equation}
\begin{equation}
\label{recurs2}
P_n = P_{n-1} - \frac{P_{n-1}}{n-1} + \frac{2}{\lambda n} + \frac{n-2}{\lambda n}.
\end{equation}Recursion (\ref{recurs1})  follows by observing that the point at which $n$ species arises creates a new interior edge from one of the $n-1$ pendant edges, hence the last term.

Recursion  (\ref{recurs2}) is more complex, but it combines the following observations: As the tree grows, from when it last has $n-1$ leaves to when it last has $n$ leaves, one of the pendant edges is selected uniformly at random from the $n-1$ pendant edges and is destroyed, becoming the new interior edge (this is the second term on the right of (\ref{recurs2})). The remaining  $n-2$ pendant edges get longer (this is the fourth term on the right of (\ref{recurs2})), and two more new pendant edges arise (the third term on the right of (\ref{recurs2})).  All these edges grow for an average of $1/\lambda n$ time (the expected time till the next event), since there are at present $n$ species and record the growth of the tree until (just before) the next speciation event.  Note that recursion  (2) simplifies to:
\begin {equation*}
P_n = P_{n-1}\left(1-\frac{1}{n-1}\right) +\frac{1}{\lambda}.
\end{equation*}
This equation, combined with the initial condition $P_2=1/2\lambda+1/2\lambda=1/\lambda$ (since the expected time of the transition from two to three leaves is $1/2\lambda$) has the closed-form solution:
$P_n = n/2\lambda.$
From this we can estimate the average expected length of a pendant edge:
$p_n= \frac{1}{n}\cdot P_n =  \frac{1}{2\lambda}$
Using this in recursion (\ref{recurs1}), along with the initial condition $I_2=0$, and the fact that there are $n-2$ interior edges,
also gives us the expected length of an interior edge, $i_n$:
$i_n= \frac{1}{n-2}\cdot I_n = \frac{1}{2\lambda}.$
In particular, $i_n=p_n$ for $n \geq 3$.  This completes the proof of Theorem~\ref{firstbasic}.
\hfill$\Box$

\subsection{Remarks} Note that the identity in Theorem~\ref{firstbasic} is under the `late sampling' scenario where the tree is observed just before the time of the next speciation event.
But if one has  $n$ leaves, and one records the $n$ pendant edge lengths  at the `earliest possible' time, namely when the $n$--th species first arises (rather than just before the $(n+1)$--st species appears) then the `correction' for the average expected length of a pendant edge will be $1/2\lambda$ - $1/n\lambda$.

Notice that if one records the pendant edge exactly half-way between these two expected times, this would give  $1/2\lambda$-$1/2n\lambda$.

However, if we observe that there are $n$ leaves in  a tree for which $t$ is unknown and we ask what is the expected time that there have been $n$ (rather than $n-1$) leaves, then this expected time
 is $1/n\lambda$ 
rather than $1/2n\lambda$, which restores our expected pendant edge estimate back to $1/2\lambda$. This follows from a result by Gernhard (\cite{ger}, Theorem 5.2, case $\mu=0$ with $k=n-1$) which studied, more generally, the distribution of times between speciation events in a birth-death tree when the age of the tree is unknown and so is assumed to have a (improper) uniform prior (see also
\cite{ger2}).  Thus, in the case where $t$ is unknown, we may assume that the expected average pendant edge length is the same as for interior edges, namely $1/2\lambda$.

Notice that in any case, the possible `corrections' all converge to $0$ as $n$ increases. This simple observation that leaf edges are the same as interior edges is behind the otherwise somewhat non-intuitive assumption made by Nee \cite{Mooers03, Nee} that one can posit a speciation event at the present when calculating diversification rates, and the fact that Pybus' gamma \cite{Pybus00} can be estimated using all weighting times, even the most recent \cite{Mooers07}.

\section{Expected average edge length in a Yule tree of given size and depth}
Let $TL_n(t)$ be the (random variable) sum of the branch lengths in a Yule tree $T$ that has depth $t$ and $n$ leaves, and let $L_n(t)$ be the expected value of $LT_n(t)$. Thus
$l_n(t): = L_n(t)/(2n-2)$ is the expected average branch length (since $T$ has $2n-2$ branches).  
\begin{theorem}
\label{mainthm}
Conditional on $n, t$ and $\lambda$, the expected value of $TL$ is given by:
$$L_n(t) = 2t+ \frac{n-2}{\lambda} (1-y(\lambda t)),$$ 
where $y(x): = \frac{xe^{-x}}{1-e^{-x}}$  is a strictly decreasing function for $x\in (0, \infty)$ with $y(0+)=1$,  $y(\infty)=0$. 
\end{theorem}
{\bf Remarks} 
Notice that (by analogy with the earlier section) we can write the expected average edge length as
 $l_n(t) = \frac{1}{2\lambda}+\delta$ where
the `correction' term $\delta = \delta(\lambda, t)$ is given by
$$\delta =  t\cdot\left(\frac{1}{n-1} - \frac{1}{(2n-2)\lambda t} - \frac{n-2}{2n-2} \cdot \frac{y(\lambda t)}{\lambda t}\right)\approx -y(\lambda t) /2\lambda $$  
where the approximation is for $n$ large.  Notice that $y(\lambda t)\rightarrow 0$  as $\lambda t \rightarrow \infty$.  Notice also  that we  can also write 
$L_n(t)= t \cdot (2+ (n-2)z(\lambda t)),$
where $z(x): =\frac{1-y(x)}{x} = 1+\frac{1}{x} - \frac{1}{1-e^{-x}}$. In  particular, we can write
$L_n(t)$ as a function of the form $t H(\lambda t)$.

\begin{figure}[ht] \begin{center}
\resizebox{4.0cm}{!}{
\input{figure1.pstex_t}
} \caption{Speciation times in a Yule tree of depth  $t$. The values $t=t_1 >t_2> t_3 > \cdots >  t_4 >  t_5=0$ measure time from the present.}
\end{center}
\label{figure1}
\end{figure}
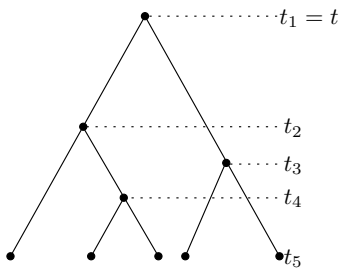

{\em Proof of Theorem~\ref{mainthm}:} Let $t_2,...,t_{n-1}$ be the (decreasing) times of the speciation events after an initial bifurcation at time $t=t_1$ in the past  --  this follows the notation of \cite{yan}, but we use $n$ here for the number of leaves, not $s$, and we write $t$ for $t_1$ (see Fig. 1 for an example with $n=5$).  The density of this vector of $t$--values conditional on $t, \lambda$ and $n$ is given by Eqn. (3) of \cite{yan} in the special case where $\rho=1, \mu=0$ (and so in the notation of that paper $p_1(t) = e^{-\lambda t}$ and $\nu_{t_1} = 1-e^{-\lambda t}$) and is given as follows:
\begin{equation}
\label{eqe0}
f({\bf t}|n, t, \lambda) = \frac{(n-2)!\lambda^{n-2}\exp(-\lambda\sum_{j=2}^{n-1}t_j)}{(1-e^{-\lambda t})^{n-2}}.
\end{equation}
Note that this is also consistent with Eqn. (5) of \cite{Nee}.
Now:
\begin{equation}
\label{eqe} 
L_n(t) = \EE[TL_n(t)] = \int_{{\bf t}} (2t+ \sum_{j=2}^{n-1}t_j) \cdot f({\bf t}|n, t, \lambda)d{\bf t},
\end{equation}
since $TL_n(t) = (2t+ \sum_{j=2}^{n-1}t_j)$, and where integration is over all tuples $(t_2,...,t_{n-1})$ for which 
$t \geq t_1 \geq t_2 \geq \cdots \geq t_{n-1} \geq 0$.
Now we can split up (\ref{eqe}) as follows:
\begin{equation}
\label{eqe1}
L_n(t)= 2t+ \int_{{\bf t}} (\sum_{j=2}^{n-1}t_j) \cdot  f({\bf t}|n, t, \lambda)d{\bf t}.
\end{equation}
From  (\ref{eqe0}),  the second term on the right-hand side of (\ref{eqe1}) is: 
\begin{equation}
\label{eqe2}
\frac{(n-2)!\lambda^{n-2}}{(1-e^{-\lambda t})^{n-2}} \cdot \int_{{\bf t}} (\sum_{j=2}^{n-1}t_j) \cdot \exp(-\lambda\sum_{j=2}^{n-1}t_j) d{\bf t}.
\end{equation}
Now we can exploit the fact that the term inside the integral sign of (\ref{eqe2}) can be written as:
\begin{equation}
\label{eqe2a}
(\sum_{j=2}^{n-1}t_j) \cdot \exp(-\lambda\sum_{j=2}^{n-1}t_j) = -\frac{d}{d\lambda}  \exp(-\lambda\sum_{j=2}^{n-1}t_j),
\end{equation}
and so, applying the Leibniz integral rule, the expression in (\ref{eqe2}) can be written as:
$$\frac{(n-2)!\lambda^{n-2}}{(1-e^{-\lambda t})^{n-2}} \cdot \left(-\frac{d}{d\lambda}\int_{{\bf t}} \exp(-\lambda\sum_{j=2}^{n-1}t_j)d{\bf t}\right).$$
Now, \begin{equation}
\label{eqe3}
\int_{{\bf t}} \exp(-\lambda\sum_{j=2}^{n-1}t_j)d{\bf t} = \frac{(1-e^{-\lambda t})^{n-2}}{(n-2)!\lambda^{n-2}}
\end{equation}
(by applying $\int_{\bf t} f({\bf t}|n, t, \lambda)d{\bf t} = 1$ to (\ref{eqe0})).
Thus, combining (\ref{eqe2}), (\ref{eqe2a}) and (\ref{eqe3}) into (\ref{eqe1}) gives:
$$L_n(t)= 2t + \frac{\lambda^{n-2}}{(1-e^{-\lambda t})^{n-2}} \cdot \left(-\frac{d}{d\lambda}\frac{(1-e^{-\lambda t})^{n-2}}{\lambda^{n-2}}\right),$$
and the result now follows by routine calculus. 
\hfill$\Box$

\subsection{Estimation of $\lambda$ from $n, t$}
Given (just) $n$ and $t$, the maximum likelihood estimate of $\lambda$,  which we denote $\lambda_{\rm ML}$, is given by:
\begin{equation}
\label{MLeq}
\lambda_{\rm ML}  = \ln\left(\frac{n}{2}\right)/t,
\end{equation}
Note that $2$ divides $n$ in this formula since we initially start with two species, and after time $t$, we observe $n$ extant species.  
Eqn. (\ref{MLeq}) can be formally verified by differentiating Eqn. (4) in \cite{Nee} with respect to $\lambda$, and solving for $\lambda$ in the resulting expression.  
With this in hand, we can now state a consequence of Theorem~\ref{mainthm}.
\begin{corollary}
\label{coro}
If we take $\lambda = \lambda_{\rm ML}$ in the expression for $\EE[TL]$ given by Theorem~\ref{mainthm} then:
$$L_n(t)= \frac{(n-2)t}{\ln(\frac{n}{2})} \mbox{  and }  \lambda_{\rm ML} =\frac{n-2}{L_n(t)}.$$
\end{corollary}
{\em Proof:} We have
$y(\lambda_{\rm ML} t)= y(\ln\left(\frac{n}{2}\right)) = \frac{2\ln(n/2)}{(n-2)}$. Thus:
$$L_n(t)= 2t+ \frac{n-2}{\lambda_{ML}}\cdot  \left(1- \frac{2\ln{(n/2)}}{(n-2)}\right) = 2t + \frac{(n-2)t}{\ln(\frac{n}{2})} - 2t = \frac{(n-2)t}{\ln(\frac{n}{2})},$$
where the second equality uses (\ref{MLeq}). This
 gives Part (i); Part  (ii) is an immediate consequence, again using (\ref{MLeq}). 

\subsection{Remarks}
Notice  that Corollary 3(i) implies that for $\lambda = \lambda_{\rm ML}$, we can express $l_n(t)$ in the familiar form of  $\frac{1}{2\lambda}$  plus a `correction term' that vanishes as $n$ grows.
More precisely, for $\lambda = \lambda_{\rm ML}$, we have:
$$l_n(t) = \frac{1}{2\lambda}(1-\frac{1}{n-1}) \approx \frac{1}{2\lambda}.$$
Nee  \cite{Nee} shows that, given a tree with branch lengths (and thereby $n, t$ and the actual value of $TL$), the maximum likelihood estimator 
of $\lambda$, which he denotes as $\hat{\lambda}$, is given by Eqn. 6 of \cite{Nee} as:
$$\hat{\lambda} = \frac{n-2}{TL}.$$
Comparing this with Corollary~\ref{coro}(ii), we see a nice concordance:  the ML estimate of $\lambda$ based on just $n$ and $t$ (i.e. $\lambda_{\rm ML}$) is exactly the same value as the ML estimate of $\lambda$ (i.e. $\hat{\lambda}$) for an actual tree whose total length $TL$ is equal to what it is expected to be under 
the Yule model for given $n$ and $t$ and $\lambda = \lambda_{\rm ML}$.
\section{Expected pendant vs. interior edge lengths as function (only) of $t$}
Let $I =I(t)$ be the expected sum of the interior edge lengths of a Yule tree that has grown for time $t$. In contrast to the previous section, the number of leaves of this tree will be regarded as an unconstrained  random variable. Similarly, let $P= P(t)$ and $L=L(t)$ be, respectively, the expected sum of the pendant (and of the total) edge lengths of a Yule tree
that has grown for time $t$. Thus, $$I(0)=P(0)=L(0)=0, \mbox{ and }  L(t) = I(t)+P(t).$$
\begin{theorem}
\label{tonly}
$$I(t) = \frac{1}{\lambda}(e^{\lambda t} + e^{-\lambda t}-2) \mbox{ and }
P(t) = \frac{1}{\lambda}(e^{\lambda t} - e^{-\lambda t}).$$
Thus, if $p(t)$ and $i(t)$ are the expected average lengths of the pendant and interior edges of a Yule tree of depth $t$, 
then the ratio $p(t)/i(t) $ converges to $1$ exponentially fast with increasing $t$. 
\end{theorem}
{\em Proof:}
From Theorem~\ref{mainthm}, $L_n(t)$ is a linear function of $n$. So,  if we regard $n$ as a random variable, rather than a given value,  then 
$L(t)$ is the expected value of $L_n(t)$ with respect to the distribution on $n$. Thus, since  $\EE[n] = 2e^{\lambda t}$,  Theorem~\ref{mainthm} gives:
$$L(t) = 2t + \frac{2e^{\lambda t} - 2}{\lambda}(1- y(\lambda t)),$$
which simplifies to:
\begin{equation}
\label{Lteq}
L(t) = \frac{2}{\lambda}(e^{\lambda t} - 1),
\end{equation}
Now, if the Yule tree has $k$ species at time $t$, then the expected sum of interior edge lengths at time $t+\delta$ is:
\begin{equation}
\label{Lteq1a}
I(t) + \delta\lambda k \cdot \frac{P(t)}{k} + o(\delta) = I(t) + \delta \lambda P(t) +o(\delta),
\end{equation} 
since $I(t)$ increases precisely if  a speciation event occurs in the interval $(t, t+\delta)$ (which has probability $\delta \lambda k +o(\delta)$)  in which case $I(t)$ increases by
the average length of pendant edges (+ $o(\delta)$), since one of the $k$ pendant edges, selected uniformly at random, becomes a new interior edge). 
Notice that the right-hand side of (\ref{Lteq1a}) is, fortunately, independent of $k$, and so:
\begin{equation}
\label{Lteq2}
\frac{dI(t)}{dt} = \lambda P(t). 
\end{equation}
Writing $P(t)=L(t)-I(t)$ in (\ref{Lteq2}) and combining this with (\ref{Lteq}) gives:
\begin{equation}
\label{Lteq2a}
\frac{dI(t)}{dt} +  \lambda I(t) = 2(e^{\lambda t}-1).
\end{equation}
This is a standard first-order linear differential equation, for which the  solution, subject to the boundary condition $I(0)=0$, is the expression for $I(t)$ 
in Theorem~\ref{tonly}. The remainder of the proof now follows easily.
\hfill$\Box$
\subsection{Remarks}
If $n$ takes its expected value $2e^{\lambda t}$, then Theorem~\ref{tonly} shows that $i(t) $ and $p(t)$ is just $\frac{1}{2\lambda}$ plus `correction terms' that
converge rapidly to $0$ with increasing $t$.
In a subsequent paper we will describe the analysis of branch lengths for the Yule model when both $n$ and $t$ are conditioned on, and when extinction is considered. The analysis in these cases is more complex, and beyond the scope of this short note.
\section{Acknowledgments}
Jonathan Davies, Walter Jetz, Tyler Kuhn, Simon Goring and Juan-Lopez Cantalapiedra sparked this problem. We also thank Tanja Stadler (n{\'e}e Gernhard) for helpful comments.

\end{document}

%% file: figure1.pstex_t
\begin{picture}(0,0)%
\epsfig{file=figure1.pstex}%
\end{picture}%
\setlength{\unitlength}{4144sp}%
\begingroup\makeatletter\ifx\SetFigFont\undefined%
\gdef\SetFigFont#1#2#3#4#5{%
  \reset@font\fontsize{#1}{#2pt}%
  \fontfamily{#3}\fontseries{#4}\fontshape{#5}%
  \selectfont}%
\fi\endgroup%
\begin{picture}(1877,1785)(239,-916)
\put(2116,-916){\makebox(0,0)[lb]{\smash{\SetFigFont{10}{12.0}{\rmdefault}{\mddefault}{\updefault}\special{ps: gsave 0 0 0 setrgbcolor}$t_5$\special{ps: grestore}}}}
\put(2087,704){\makebox(0,0)[lb]{\smash{\SetFigFont{10}{12.0}{\rmdefault}{\mddefault}{\updefault}\special{ps: gsave 0 0 0 setrgbcolor}$t_1=t$\special{ps: grestore}}}}
\put(2116,-511){\makebox(0,0)[lb]{\smash{\SetFigFont{10}{12.0}{\rmdefault}{\mddefault}{\updefault}\special{ps: gsave 0 0 0 setrgbcolor}$t_4$\special{ps: grestore}}}}
\put(2116,-286){\makebox(0,0)[lb]{\smash{\SetFigFont{10}{12.0}{\rmdefault}{\mddefault}{\updefault}\special{ps: gsave 0 0 0 setrgbcolor}$t_3$\special{ps: grestore}}}}
\put(2116,-35){\makebox(0,0)[lb]{\smash{\SetFigFont{10}{12.0}{\rmdefault}{\mddefault}{\updefault}\special{ps: gsave 0 0 0 setrgbcolor}$t_2$\special{ps: grestore}}}}
\end{picture}